\documentclass[fleqn]{article}
\usepackage{bm,amsfonts,amssymb,amsmath,graphicx,enumerate}

\begin{document}
\title{Conformal flatness, non-Abelian Kaluza-Klein \\ and the Quaternions}
\author{Paolo Maraner, Jiannis K.\ Pachos}
\maketitle
\begin{abstract}
\noindent
Maximally symmetric manifolds with holonomy in the unitary quaternionic group $Sp(d/4)$
emerge from the non-Abelian Kaluza-Klein reduction of conformally flat spaces.
Thus, all special manifolds with constant properly `holonomy-related' sectional
curvature, are in natural correspondence with conformally flat, possibly non-Abelian, Kaluza-Klein spaces.
\end{abstract}

\subsection*{Introduction}\label{Intro}

Maximally symmetric real manifolds, like deSitter and anti-deSitter spacetimes,
are conformally flat. Maximally symmetric manifolds with special  holonomy are not.
When holonomies lie in the unitary group $U(d/2)$, however, the relation between maximal
symmetry and conformal flatness is not completely lost. The investigation of the Kaluza-Klein reduction
(codimension 1) of geometric tensors characterizing conformal flatness initiated by R.\ Jackiw and
collaborators \cite{Guralinik&Iorio&Jackiw&Pi03,Grumiller&Jackiw06,Jackiw07,Grumiller&Jackiw07},
has shown that complex manifolds with constant holomorphic sectional curvature are naturally associated
with conformally flat Kaluza-Klein spacetimes \cite{Maraner&Pachos09}.
Here we extend the investigation to the non-Abelian Kaluza-Klein reduction from $D$ to $d=D-c$
dimensions (codimension $c$) of conformal tensors, showing that maximally symmetric manifolds with holonomies
in the unitary quaternionic group $Sp(d/4)$, emerge from the dimensional reduction of conformally flat spaces.
Thus, all special manifolds with constant properly `holonomy-related' sectional curvature, are in natural
correspondence with higher dimensional conformally flat, possibly non-Abelian, Kaluza-Klein spaces.\\
Our discussion proceeds as follows. In the next section we briefly review quaternionic $Sp(d/4)$ structures
on manifolds, with emphasis on the maximally symmetric case. Next, we turn to the non-Abelian dimensional
reduction of the tensors characterizing conformal flatness in three and higher dimension and to the equations
obtained by imposing their vanishing. The following two sections are devoted to the construction of non-trivial
solutions of these equations. These turn out to be non-Abelian Kaluza-Klein spaces with internal 3-dimensional
spheres and external maximally symmetric quaternionic K\"{a}hler spaces. The last section contains our conclusions,
while two Appendices display general non-Abelian reduction formulas for the Cotton and Weyl tensors.

\subsection*{Quaternionic K\"{a}hler manifolds}\label{Q}

In the last few decades there has been an increasing interest
in manifolds with special holonomy \cite{Bryant93}.
Besides \emph{complex}  manifolds, with holonomy in $U(d/2)$
\cite{Yano64,Barros&Romero82,Cruceanu&Fortuny&Gadea96},
also \emph{quaternionic} manifolds, with holonomy in $Sp(d/4)$
\cite{Alekseevskii68,Ishihara74,Salomon82,Garcia-Rio&Matsushita&Vazquez-Lorenzo01},
have been introduced and investigated. Among these, maximally
symmetric spaces, i.e. spaces of constant properly `holonomy-related'
sectional curvature, are regarded as the simplest important classes (see also
\cite{Gadea&AmilibiaMontesinos89,Gadea&MunozMasque92,Perez&Santos82,Blazic92}).

In analogy with the complex case, the holonomy restriction to the unitary quaternionic group
$Sp(d/4)$ is obtained by imposing
an appropriate Clifford-K\"{a}hler structure on the base manifold \cite{Ishihara74}. Namely, one introduces a metric $g_{\mu\nu}$
and three complex structures
$J_{\mu}^{{\sf a}\nu}$, ${\sf a}=1,2,3$,
closing the quaternionic algebra
\begin{subequations}\label{quaternionic algebra}
\begin{equation}
J_{\mu}^{{\sf a}\kappa}J_{\kappa}^{{\sf b}\nu}+J_{\mu}^{{\sf b}\kappa}J_{\kappa}^{{\sf a}\nu}=
-2\,\delta^{{\sf a}{\sf b}}\delta_\mu^\nu,
\label{Clifford}
\end{equation}
\begin{equation}
J_{\mu}^{{\sf a}\kappa}J_{\kappa}^{{\sf b}\nu}-J_{\mu}^{{\sf b}\kappa}J_{\kappa}^{{\sf a}\nu}=
2\,{\varepsilon^{{\sf a}{\sf b}}}_{\sf c}J_{\mu}^{{\sf c}\nu},
\label{so-algebra}
\end{equation}\end{subequations}
where ${\varepsilon_{{\sf a}{\sf b}{\sf c}}}$ is
the completely antisymmetric Levi-Civita symbol. The compatibility
between the metric and the quaternionic structures is obtained by requiring the $J_{\mu}^{{\sf a}\nu}$
to be isometries
\begin{subequations}\label{HK}
\begin{equation}
J_{\mu}^{{\sf a}\kappa}J_{\nu}^{{\sf a}\lambda}g_{\kappa\lambda}=g_{\mu\nu}
\label{Hermitian}
\end{equation}
(no sum over ${\sf a}$), parallel with respect to the Levi-Civita connection $\nabla_\kappa$
associated to $g_{\mu\nu}$
\begin{equation}
\nabla_\kappa J_{\mu}^{{\sf a}\nu}={\varepsilon_{{\sf b}{\sf c}}}^{\sf a}\theta_\kappa^{\sf b}J_{\mu}^{{\sf c}\nu},
\label{Kaehler}
\end{equation}\end{subequations}
where $\theta^{\sf a}_\mu$ are appropriate 1-forms. This last equation is a generalization of the K\"ahler
condition and the resulting geometries are called \emph{quaternionic K\"ahler geometries}.\\
Still in complete analogy with the complex case, one defines quaternionic sectional curvatures along
$J_{\mu}^{{\sf a}\nu}$-invariant  subspaces and finds that the manifolds of constant
quaternionic  sectional curvature $k$, are characterized by a Riemann curvature tensor of the form
\begin{subequations}\label{constant k}
\begin{equation}\begin{array}{l}
R_{\mu\nu\kappa\lambda}=\frac{1}{4}k
\left(g_{\mu\lambda}g_{\kappa\nu}-g_{\mu\kappa}g_{\lambda\nu}
+ J_{{\sf a}\mu\lambda}J^{\sf a}_{\nu\kappa}- J_{{\sf a}\mu\kappa}J^{\sf a}_{\nu\lambda}
-2 J_{{\sf a}\mu\nu}J^{\sf a}_{\kappa\lambda}\right),
\label{SpaceFormCurvature}
\end{array}\end{equation}
with the 2-forms $J_{\mu\nu}^{\sf a}=J_{\mu}^{{\sf a}\kappa}g_{\kappa\nu}$, related to the 1-forms $\theta^{\sf a}_\mu $ by
\begin{equation}
J_{\mu\nu}^{\sf a}=\frac{1}{k}\left(\partial_\mu\theta^{\sf a}_\nu-\partial_\nu\theta^{\sf a}_\mu
-{\varepsilon_{{\sf b}{\sf c}}}^{\sf a}\theta^{\sf b}_\mu \theta^{\sf c}_\nu\right).
\label{J as a Gauge Field}
\end{equation}\end{subequations}
In what follows we show that precisely this structure
emerges form the non-Abelian Kaluza-Klein reduction of conformally flat spaces.

\subsection*{Dimensional reduction of conformal tensors}\label{DR}
A space is said to be {\em conformally flat} if a coordinate
transformation exists, making its metric tensor proportional to a
locally flat metric. All one- and two-dimensional spaces are
conformally flat. In three dimensions conformal flatness is
characterized by the vanishing of the Cotton tensor. In
dimensions greater than three  Cotton flatness reduces to a
necessary but non-sufficient condition and the role of probing
conformal flatness is taken by the vanishing of the Weyl tensor.
We consider a $D$-dimensional pseudo-Riemannian space
with $D\geq3$. The metric tensor
${\mathbf g}_{IJ}({\mathbf x})$ is allowed to carry arbitrary
signature, while the corresponding Riemann, Ricci and scalar
curvatures are respectively defined as
${{\mathbf R}_{IJK}}^L=\partial_I{{\mathbf\Gamma}_{JK}}^L-...$,
${\mathbf R}_{IJ}={{\mathbf R}_{IKJ}}^K$ and
${\mathbf R}={{\mathbf R}_{I}}^I$.\footnote{(Higher)
$D$-dimensional quantities are denoted in boldface characters with
capital Latin indices $I,J,...=1,...,D$.}
The Cotton tensor ${\mathbf C}_{IJK}$ and the Weyl tensor ${\mathbf C}_{IJKL}$
are then conveniently written in terms of the Schouten
tensor ${\mathbf S}_{IJ}={\mathbf R}_{IJ}-\frac{1}{2(D-1)}
{\mathbf R}{\mathbf g}_{IJ}$ as\footnote{Here and in what follows,
square and round brackets respectively denote anti-symmetrization $\textbf{t}_{[IJ]}=(\textbf{t}_{IJ}-\textbf{t}_{JI})/2$
and symmetrization $\textbf{t}_{(IJ)}=(\textbf{t}_{IJ}+\textbf{t}_{JI})/2$.}
\begin{equation}
{\mathbf C}_{IJK}\equiv2\bm{\nabla}_{[K}{\mathbf S}_{J]I},\label{Cotton}
\end{equation}
\begin{equation}
{\mathbf C}_{IJKL}\equiv{\mathbf R}_{IJKL}-\frac{2}{D-2}\left(
{\mathbf g}_{I[K}{\mathbf S}_{L]J} - {\mathbf g}_{J[K}{\mathbf
S}_{L]I}\right).\label{Weyl}
\end{equation}
 The Cotton tensor is antisymmetric in the last two indices by
construction and is traceless in the first two because of the
Bianchi identities. The Weyl tensor shares the same symmetries of
the Riemann tensor and it is further vanishing under contraction
of any pair of indices. In dimension grater than three, Cotton
and Weyl  tensors are related by the identity
\begin{equation}
(D-3) {\mathbf C}_{IJK}=(D-2)\bm{\nabla}_{L} {{\mathbf
C}_{IJK}}^L.\label{Cotton vs Weyl}
\end{equation}
The vanishing of the latter implies the vanishing of the former,
but the vice versa does not hold.

We reduce geometry to arbitrary dimensions $d$ and codimensions $c\geq2$, $d+c=D$,
by a standard \emph{non-Abelian Kaluza-Klein Ansatz} \cite{KK}
\begin{equation}\label{KKmetric}
{\mathbf g}_{IJ}= \left(
\begin{array}{cc}
g_{\mu\nu}+A_\mu^{\sf a}A_\nu^{\sf b}K_{\sf a}^kK_{\sf
b}^l\kappa_{kl}
& A_\mu^{\sf a}K_{\sf a}^k\kappa_{kj}\\
\kappa_{il}K_{\sf a}^l A_\nu^{\sf a}& \kappa_{ij}
\end{array}
\right).
\end{equation}
The lower dimensional tensors $g_{\mu\nu}(x)$, $\kappa_{ij}(y)$ are identified with
the external $d$-dimensional  and internal $c$-dimensional  metrics, respectively. 
The Killing vector fields $K_{\sf a}^k(y)$ span the internal isometry algebra ${\cal I}$,
$[K_{\sf a},K_{\sf b}]^i={c_{\sf ab}}^{\sf c}K^i_{\sf c}$, and
$A_\mu^{\sf a}(x)$ is an external vector field
taking values on ${\cal I}$.\footnote{(Lower) external
$d$- and internal $c$-dimensional indices are respectively denoted by small
Greek and Latin characters, $\mu,\nu=...=1,...,d$ and $i,j,...=1,...,c$. Internal/external
coordinates
are respectively denoted by  $x^\mu$, $y^i$, while the corresponding curvatures
by $R_{\mu\nu\kappa\lambda}$, $R_{\mu\nu}$, $R^\mathrm{ex}$ and $R_{ijkl}$, $R_{ij}$,
$R^\mathrm{in}$. The indices ${\sf a}, {\sf b}, ...$ range over the dimension $n$
of the internal isometry algebra ${\cal I}$.} The non-Abelian Kaluza-Klein Ansatz is covariant under the
coordinate transformations generated by $\delta x^\mu=\xi^\mu(x)$, $\delta y^i=\varepsilon^{\sf a}(x)K_{\sf a}^i(y)$.
The former transformations are identified with external space-time diffeomorphisms. The latter provide the gauge transformation of the gauge potential $A_\mu^{\sf a}$.
The corresponding gauge curvature is obtained as
$F_{\mu\nu}^{\sf a}=\partial_\mu A_\nu^{\sf a} -\partial_\nu
A_\mu^{\sf a} - c_{\sf bc}^{\sf\ \ a}A_\mu^{\sf b}A_\nu^{\sf c}$.
The external covariant derivative $\nabla_\mu$
associated to $g_{\mu\nu}$, is not covariant under the second group of
transformations when acting on
internal scalars/tensors. It has to be replaced by
\[
\hat\nabla_\mu\equiv\nabla_\mu-{\mathcal L}_{A_\mu},
\]
with ${\mathcal L}_{A_\mu}$ the Lie derivative with respect to the
internal vector field $A^i_\mu\equiv K_{\sf a}^iA_\mu^{\sf a}$. When
acting on  $F^i_{\mu\nu}\equiv K_{\sf a}^i F^{\sf
a}_{\mu\nu}$, $\hat\nabla_\mu$ reproduces the familiar gauge covariant derivative
\begin{equation}\label{DF}
\hat\nabla_\kappa F^{i}_{\mu\nu}= K_{\sf a}^i
(\nabla_\kappa F^{\sf a}_{\mu\nu}-{c_{\sf bc}}^{\sf a}A^{\sf
b}_{\kappa} F^{\sf c}_{\mu\nu}).
\end{equation}
It is standard matter to obtain reduction formulas for
the higher dimensional curvatures ${{\mathbf R}_{IJKL}}$, ${{\mathbf R}_{IJ}}$, ${{\mathbf R}}$, in terms of the lower dimensional ones
$R_{\mu\nu\kappa\lambda}$, $R_{ijkl}$, ..., and of the gauge fields $A_\mu^{\sf a}$, $F^{\sf a}_{\mu\nu}$.
Substituting these in the higher dimensional Einstein equations produces the standard $d$ dimensional Einstein-Yang-Mills theory.
Our interest, instead, is in analogue reduction formulas for the higher dimensional
conformal tensors and in the equations obtained by imposing
their vanishing. These describe the `immersion' of $d$ dimensional
space-time into a $D=d+c$ dimensional conformally flat space, also reproducing
the equations of motion of a non-standard $d$ dimensional
Einstein-Yang-Mills theory. General reduction formulas for the Cotton and the Weyl tensors
are respectively presented in Appendices \ref{A} and \ref{B}. Here, we proceed to the analysis
of the equations for conformal flatness.

\subsection*{Equations for conformal flatness}\label{Equations}
Demanding conformal flatness in dimensions greater than three,
amounts to set all  components of the Weyl tensor equal to
zero.\footnote{In three dimensions conformal flatness requires the Cotton
tensor to vanish identically. The  only possibility of having $c>1$ implies
a lineal spacetime. The external scalar curvature and the gauge curvature
are consequently equal to zero. The formulas of Appendix \ref{A}
show that the only non-identically vanishing Cotton components
are (\ref{Ci}) and (\ref{Cijk}). Setting them equal to zero implies the
single equation $\partial_k R^\mathrm{in}=0$.
The structure of $D=3$, $c=2$ conformally flat Kaluza-Klein spaces
is trivial: vanishing $F^{\sf a}_{\mu\nu}$, pure gauge $A^{\sf
a}_{\mu}$, vanishing external scalar curvature $R^\mathrm{ex}$ and constant
internal scalar curvature $R^\mathrm{in}$.}
From the dimensional reduction formulas of Appendix \ref{B} we obtain the following set of equations
\begin{subequations}\label{Weyl=0}
\begin{equation}\begin{array}{l}
C_{\mu\nu\kappa\lambda}+
\frac{1}{2}\left(F_{k\mu\nu}F^k_{\kappa\lambda}-
F_{k\mu[\kappa}F^k_{\lambda]\nu}\right)
-\frac{3}{2(d-2)}\left(g_{\mu[\kappa} T_{\lambda]\nu}-
g_{\nu[\kappa}T_{\lambda]\mu}\right)=0 \label{GJ1}
\end{array}\end{equation}
\begin{equation}\begin{array}{l}
R_{\mu\nu}-\frac{1}{d} R^\mathrm{ex}g_{\mu\nu}=-
\frac{d+3c-2}{4c} \left(F^2_{\mu\nu}
-\frac{1}{d}F^2g_{\mu\nu}\right)\label{GJ2}
\end{array}\end{equation}
\begin{equation}\begin{array}{l}
\hat\nabla_\kappa F_{i\mu\nu}=0
\label{DF=0}
\end{array}\end{equation}
\begin{equation}\begin{array}{l}
C_{ijkl}=0 \label{GJ1i}
\end{array}\end{equation}
\begin{equation}\begin{array}{l}
R_{ij}-\frac{1}{c} R^\mathrm{in}\kappa_{ij}=-\frac{c-2}{4d}
\left(F^2_{ij} -\frac{1}{c}F^2\kappa_{ij}\right)\label{GJ2i}
\end{array}\end{equation}
\begin{equation}\begin{array}{l}
F_{(i|\mu\kappa}{F_{j)\nu}}^\kappa=\frac{1}{c}F^2_{\mu\nu} \kappa_{ij}+\frac{1}{d}g_{\mu\nu}
F^2_{ij}-\frac{1}{cd}F^2
g_{\mu\nu}\kappa_{ij}\label{FFs}
\end{array}\end{equation}
\begin{equation}\begin{array}{l}
F_{[i|\mu\kappa}{F_{j]\nu}}^\kappa=2\nabla_{i}F_{j\mu\nu} \label{FFa}
\end{array}\end{equation}
\begin{equation}\begin{array}{l}
c(c-1)R^\mathrm{ex}+d(d-1)R^\mathrm{in}+
\frac{(c-1)(2d+3c-2)}{4}F^2=0\label{C=0}
\end{array}
\end{equation}\\
\end{subequations}
with
\begin{equation}
F^2_{\mu\nu}=F^k_{\mu\kappa}{F_{k\nu}}^\kappa,\hskip0.3cm
F^2_{ij}=F_{i\mu\nu}F_j^{\mu\nu},\hskip0.3cm
F^2=F^i_{\mu\nu}F_i^{\mu\nu}
\end{equation}
and $\,T_{\mu\nu}=F^2_{\mu\nu}-\frac{1}{2(d-1)}F^2g_{\mu\nu}$.
The first, second and third equations are the higher co-dimensional generalization of the equations
obtained by Grumiller and Jackiw in codimension one (see (17a,b,c) in Ref.~\cite{Grumiller&Jackiw06}).
The only remarkable difference is that the whole gauge field is required to be covariantly constant
and not only its traceless part. The fourth and fifth equations
appear as the internal counterpart of the first and second ones. The sixth and seventh equations fix
the internal symmetric and antisymmetric part of the gauge field contraction $F_{i\mu\kappa}{F_{j\nu}}^\kappa$.
Eventually, the eighth equation provides a relation among the gauge 
 squared modulus and the external and
internal curvatures. All equations but (\ref{DF=0}), are traceless in all paired internal/external indices.

Given the complexity of the problem, it is useful to establish
integrability conditions for (\ref{Weyl=0}). In view of (\ref{Cotton vs Weyl}),
these are obtained by imposing the vanishing of the Cotton tensor.
From the formulas of Appendix \ref{A} and by taking into account (\ref{GJ2}),
(\ref{DF=0}), (\ref{GJ1i}) and (\ref{C=0}), we find that the only non-trivially satisfied conditions are
\begin{subequations}\label{Cotton=0}
\begin{equation}\begin{array}{l}
 F_{k\mu}^\kappa R_{\kappa\nu}+F^l_{\mu\nu}R_{lk}
 +\frac{1}{2}F_{k\mu}^\kappa F^2_{\kappa\nu}
 -\frac{1}{4}F^l_{\mu\nu}F^2_{lk}=0,\label{EqCottonC}
\end{array}\end{equation}
\begin{equation}\begin{array}{l}
\partial_i \left(R^\mathrm{in}- \frac{3d+4c-4}{4d}F^2\right)=0.\label{EqCottonF}
\end{array}\end{equation}\\ \end{subequations}
Equations (\ref{Weyl=0}), together with the integrability conditions (\ref{Cotton=0}), are solved by
vanishing $F^{\sf a}_{\mu\nu}$, pure gauge $A^{\sf a}_{\mu}$ and maximally symmetric external and internal
spaces with (constant) scalar curvatures related by
\begin{equation}
c(c-1)R^\mathrm{ex}+d(d-1)R^\mathrm{in}=0.
\label{RexRin}
\end{equation}
The question is whether less trivial solutions exist, carrying non-vanishing non-Abelian gauge configurations
endowing the external space with interesting geometrical structures, like in codimension one \cite{Maraner&Pachos09}.

\subsection*{Curvatures}\label{Curvatures}
We start by analyzing the constraints imposed by equations (\ref{Weyl=0}) on curvatures.
First, we prove that the scalars $R^\mathrm{ex}$, $R^\mathrm{in}$ and $F^2$ have to be constant.
By considering the internal derivative of (\ref{C=0}) together with
(\ref{EqCottonF}), one obtains the simultaneous linear equations
\begin{equation}\left\{
\begin{array}{l}
4d(d-1)\,\partial_iR^\mathrm{in}+
{(c-1)(2d+3c-2)}\,\partial_iF^2=0\\
4d\,\partial_iR^\mathrm{in}-
{(3d+4c-4)}\,\partial_iF^2=0
\end{array},\right.
\end{equation}
that impose the vanishing of $\partial_iR^\mathrm{in}$ and $\partial_iF^2$ for $D\neq1$ and $D\neq2$.
It follows that the internal curvature $R^\mathrm{in}$ is constant, while
the squared modulus of the gauge curvature $F^2$ is only allowed to depend
on external coordinates. On the other hand, (\ref{DF=0}) implies
\begin{equation}
\partial_\kappa F^2=\hat\nabla_\kappa F^i_{\mu\nu}F_i^{\mu\nu}=(\hat\nabla_\kappa F^i_{\mu\nu})F_i^{\mu\nu}+
F^i_{\mu\nu}(\hat\nabla_\kappa F_i^{\mu\nu})=0,\nonumber
\end{equation}
requiring $F^2$ not to depend on external coordinates, hence its constancy. Eventually, (\ref{C=0}) requires
the constancy of the external curvature $R^\mathrm{ex}$, fixing its value to
\begin{equation}\begin{array}{l} R^\mathrm{ex}=-\frac{d(d-1)}{c(c-1)}R^\mathrm{in}-\frac{2d+3c-2}{4c}F^2.
\label{EqWeylH'}
\end{array}\end{equation}
Next, we consider the curvature tensors.
Along internal directions, equation (\ref{GJ1i}) together with the constancy of the internal scalar curvature
$R^\mathrm{in}$, immediately implies that the internal space is maximally symmetric.
The internal Riemann and Ricci curvatures rewritten in terms of the constant $R^\mathrm{in}$ and of
the internal metric $\kappa_{ij}$ are then given by
\begin{subequations}\label{InCurvatures}
\begin{equation}\begin{array}{l}
R_{ijkl}=\frac{1}{c(c-1)}R^\mathrm{in}\left(\kappa_{ik}\kappa_{lj}-\kappa_{il}\kappa_{kj}\right),
\label{RiemannIn}
\end{array}\end{equation}
\begin{equation}\begin{array}{l}
R_{ij}=\frac{1}{c}R^\mathrm{in}\kappa_{ij}.
\label{RicciIn}
\end{array}\end{equation}\end{subequations}\\
The internal space is isomorphic  to the pseudo-Euclidean space $\mathbb{R}^c_s$ for $R^\mathrm{in}=0$,
to the pseudo-projective space $\mathbb{R}P^c_s$ (the pseudo-sphere $S^c_s$) for $R^\mathrm{in}>0$ and to
the pseudo-hyperbolic space $\mathbb{R}H^c_s$ for $R^\mathrm{in}<0$, $s=0,...,c$ (see e.g. \S8 of Ref.~\cite{O'Neill83}).
In particular, the internal space is Einstein, so that for $c>2$ equation (\ref{GJ2i}) imposes the vanishing of
the traceless part of the symmetric tensor $F^2_{ij}$.\\
Along external directions, we solve equation (\ref{GJ1}) by the substitution
\begin{equation}\begin{array}{l}
R_{\mu\nu\kappa\lambda}=\texttt{r}_{\mu\nu\kappa\lambda}-\frac{1}{2}\left(F_{k\mu\nu}F^k_{\kappa\lambda}-
F_{k\mu[\kappa}F^k_{\lambda]\nu}\right),\label{Rmnkl}
\end{array}\end{equation}
with $\texttt{r}_{\mu\nu\kappa\lambda}$  a tensor sharing the algebraic symmetries of
a Riemann tensor and further satisfying the conditions
\begin{equation}\begin{array}{l}
\texttt{r}_{\mu\nu\kappa\lambda}
-\frac{2}{d-2}\left(g_{\mu[\kappa}\texttt{r}_{\lambda]\nu}
+g_{\nu[\kappa}\texttt{r}_{\lambda]\mu}\right)-
\frac{2}{(d-1)(d-2)}\texttt{r}g_{\mu[\kappa}g_{\lambda]\nu}=0,
\end{array}\end{equation}
with $\texttt{r}_{\mu\nu}={\texttt{r}_{\mu\kappa\nu}}^\kappa$ and
$\texttt{r}={\texttt{r}_\mu}^\mu$.
These are $\frac{1}{12}d(d+1)(d+2)(d-3)$
simultaneous linear equations in $\frac{1}{12}d^2(d^2-1)$ variables, with
coefficients only depending on the external spacetime metric $g_{\mu\nu}$. The
general solution depends on $\frac{1}{2}d(d+1)$ arbitrary functions and is obtained as
\begin{equation}\begin{array}{l}
\texttt{r}_{\mu\nu\kappa\lambda}=2\left(g_{\mu[\lambda}\rho_{\kappa]\nu}-
g_{\nu[\lambda}\rho_{\kappa]\mu}\right)+2\rho
g_{\mu[\lambda}g_{\kappa]\nu}\label{rmn},
\end{array}\end{equation}
with  $\rho_{\mu\nu}$ a traceless symmetric tensor and $\rho$ a
scalar.  The quantities $\rho_{\mu\nu}$ and $\rho$ are
determined by equations (\ref{GJ2}) and (\ref{EqWeylH'}) as follows.
From (\ref{Rmnkl}) and  (\ref{rmn}) we obtain
$R_{\mu\nu}=(d-1)\rho g_{\mu\nu}+(d-2)\rho_{\mu\nu}-\frac{3}{4}
F_{\mu\kappa}{F_\nu}^\kappa$ and $R^\mathrm{ex}=d(d-1)\rho-\frac{3}{4}F^2$, which
substituted back in (\ref{GJ2}) and (\ref{EqWeylH'}) yields
\begin{equation}\begin{array}{l}
\rho_{\mu\nu}=-\frac{1}{4c} \left(F^2_{\mu\nu}
-\frac{1}{d}F^2g_{\mu\nu}\right),\hskip0.4cm\rho=-\frac{1}{c(c-1)}R^\mathrm{in}-
\frac{1}{2cd}F^2g_{\mu\nu}.\label{rhomn&rho}
\end{array}\end{equation}
Eventually, by substituting (\ref{rhomn&rho}) back in (\ref{rmn}) and this in (\ref{Rmnkl}),
the external Riemann and Ricci tensors are obtained in terms of the external metric
$g_{\mu\nu}$, the internal metric $\kappa_{ij}$, the gauge field $F^k_{\mu\nu}$ and the
constant $R^\mathrm{in}$ as
\begin{subequations}\label{ExCurvatures}
\begin{equation}\begin{array}{rl}
R_{\mu\nu\kappa\lambda}=&-\frac{1}{c(c-1)}R^\mathrm{in}\left(g_{\mu\kappa}g_{\lambda\nu}-g_{\mu\lambda}g_{\kappa\nu}\right)\\
&-\frac{1}{2c}\left(g_{\mu[\kappa}F^2_{\lambda]\nu}-g_{\nu[\kappa}F^2_{\lambda]\mu}\right)
-\frac{1}{2}\left(F_{k\mu\nu}F^k_{\kappa\lambda}-F_{k\mu[\kappa}F^k_{\lambda]\nu}\right),
\label{RiemannEx}
\end{array}\end{equation}
\begin{equation}\begin{array}{l}
R_{\mu\nu}=-\frac{d-1}{c(c-1)}R^\mathrm{in}g_{\mu\nu}-\frac{d+3c-2}{4c}F^2_{\mu\nu}-\frac{1}{4c}F^2g_{\mu\nu}.
\label{RicciEx}
\end{array}\end{equation}\end{subequations}
Equation (\ref{DF=0}), together with the constancy of $R^\mathrm{in}$, guarantee that the Bianchi integrability conditions are satisfied.
\vskip0.2cm
Further information on curvatures can be obtained from the integrability condition (\ref{EqCottonC}).
While the analysis of this equation is possible in general, for simplicity of presentation we restrict
to $c>2$, where equation (\ref{GJ2i}) imposes the vanishing of the traceless part of $F^2_{ij}$.
By inserting (\ref{RicciIn}) and (\ref{RicciEx}) in (\ref{EqCottonC}) we obtain
\[\begin{array}{l}
{F_{i\lambda}}^{\kappa}{F^{2\nu}_\kappa}=-\frac{4}{c-1}R^\mathrm{in}{F_{i\lambda}}^{\nu}.
\end{array}\]
Contraction  with $F_\mu^{i\lambda}$ yields
\begin{equation}\label{FF=F}
\begin{array}{l}
F_\mu^{2\kappa}F_\kappa^{2\nu}=-\frac{4}{c-1}R^\mathrm{in} F_\mu^{2\nu}.
\end{array}
\end{equation}
Restricting attention to solutions with non-degenerate $F_\mu^{2\nu}$, equation (\ref{FF=F}) implies
the proportionality between $F^2_{\mu\nu}$ and the external metric $g_{\mu\nu}$
\begin{equation}\begin{array}{l}
F^2_{\mu\nu}=-\frac{4}{c-1}R^\mathrm{in}g_{\mu\nu}.
\end{array}\label{F2munu}
\end{equation}
One more contraction yields a relation expressing the gauge field modulus in terms of the internal curvature
\begin{equation}\begin{array}{l}
F^2=-\frac{4d}{c-1}R^\mathrm{in}.
\label{F^2 vs R^in}
\end{array}\end{equation}
This last two equations, together with (\ref{GJ2}) imply the external space is Einstein.
By inserting them in (\ref{ExCurvatures}) we eventually obtain the external Riemann and Ricci curvatures as
\begin{subequations}\label{ExCurvatures'}
\begin{equation}\begin{array}{l}
R_{\mu\nu\kappa\lambda}=-\frac{1}{2cd}F^2 g_{\mu[\kappa}g_{\lambda]\nu}
-\frac{1}{2}\left(F_{k\mu\nu}F^k_{\kappa\lambda}-F_{k\mu[\kappa}F^k_{\lambda]\nu}\right),
\label{RiemannEx'}
\end{array}\end{equation}
\begin{equation}\begin{array}{l}
R_{\mu\nu}=-\frac{d+3c-1}{4cd}F^2g_{\mu\nu}.
\label{RicciEx'}
\end{array}\end{equation}\end{subequations}

\subsection*{Absolute parallelism and Clifford structures}
Next, we turn our attention to equations (\ref{FFs}) and (\ref{FFa}). Contracting both equations
with $K_{\sf a}^iK_{\sf b}^j$, taking into account (\ref{F2munu}) and the  commutation relations
$\left[K_{\sf a},K_{\sf b}\right]^i={c_{{\sf a}{\sf c}}}^{\sf c}K_{\sf c}^i$,  we
respectively rewrite them as
\begin{subequations}\label{FF'}
\begin{equation}
{\sf g}_{{\sf a}{\sf c}}{\sf g}_{{\sf b}{\sf d}}
\left(F^{{\sf c}\kappa}_{\mu}F^{{\sf d}\nu}_{\kappa}+F^{{\sf d}\kappa}_{\mu}F^{{\sf c}\nu}_{\kappa}\right)=
-\frac{2}{d}\,{\sf g}_{{\sf a}{\sf c}}{\sf g}_{{\sf b}{\sf d}}F^{\sf c}_{\kappa\lambda}F^{{\sf d}\kappa\lambda}\delta_\mu^\nu,
\label{FFs'}
\end{equation}
\begin{equation}
{\sf g}_{{\sf a}{\sf c}}{\sf g}_{{\sf b}{\sf d}}
\left(F^{{\sf c}\kappa}_{\mu}F^{{\sf d}\nu}_{\kappa}-F^{{\sf d}\kappa}_{\mu}F^{{\sf c}\nu}_{\kappa}\right)=
2\left(K_{\sf a}^i\partial_i{\sf g}_{{\sf b}{\sf c}}
-K_{\sf b}^i\partial_i{\sf g}_{{\sf a}{\sf c}}
-{c_{{\sf a}{\sf b}}}^{\sf c}{\sf g}_{{\sf c}{\sf d}}\right)F^{{\sf c}\nu}_{\mu},
\label{FFa'}
\end{equation}
\end{subequations}
where ${\sf g}_{{\sf a}{\sf b}}=K_{\sf a}^iK_{\sf b}^j\kappa_{ij}$
is the matrix implicitly appearing in all gauge field internal contractions.
The similarity of equations (\ref{FF'}) with (\ref{quaternionic algebra})
for constant ${\sf g}_{{\sf a}{\sf b}}$ is striking.
We therefore restrict attention to solution with constant ${\sf g}_{{\sf a}{\sf b}}$.\footnote{We suspect these
to be the only possible solutions, but we could not prove the statement.}
This assumption has strong implications. Proceeding by a Gram-Schmidt process, applied
to $K_1^i, ..., K_n^i$ at one point, we obtain orthonormal Killing vector fields globally.
On the one hand, the number of these can not exceed the internal dimension $c$.
On the other hand, for $c>2$ equation (\ref{GJ2i}) requires
\[K_{{\sf a}i}F^{\sf a}_{\mu\nu}F^{{\sf b}\mu\nu}K_{{\sf b}j}=\frac{1}{c}F^2\kappa_{ij}.\]
The right hand side term of this equation is a non-singular $c \times c$ matrix. Hence, the
$c\times n$, $n \times n$ and $n\times c$ matrices appearing  in the left hand side term, should have
at least rank $c$. This implies the presence of at least $c$ linearly independent Killing vector fields.
Hence, the number of orthonormal Killing vector fields $K_{\sf a}^i$ corresponding to non-vanishing
$F^{\sf a}_{\mu\nu}$ has to be exactly equal to $c$.
That is, the internal space has to support a global Killing parallelization.
The only manifolds of constant curvature admitting a Killing parallelization are isomorphic
to the $c$-dimensional pseudo-Euclidean space $\mathbb{R}^c_s$, $s=0,...,c$, the
1-dimensional sphere $S^1$, the 3-dimensional (pseudo-)spheres $S^3$, $S^3_1$,
the 7-dimensional (pseudo-)spheres $S^7$, $S^7_3$.
The special dimensions of the (pseudo-)spheres depend essentially on the existence of a multiplication in
$\mathbb{R}^2$ (complex numbers $\mathbb{C}$), $\mathbb{R}^4$ (quaternions $\mathbb{H}$),
and $\mathbb{R}^8$ (octonions $\mathbb{O}$) \cite{D'Atri&Nickerson68,Wolf72}. \\
The isometry algebras of $\mathbb{R}^c_s$ and $S^1$ are Abelian, taking the analysis back to \cite{Maraner&Pachos09}.\\
In the non-Abelian case, the internal space is isomorphic to  $S^3$, $S^3_1$, $S^7$  or $S^7_3$.
For simplicity of the presentation we restrict our analysis to $S^3$ and $S^7$.
The remaining two cases are treated along  the very same lines.
After the orthonormalization process, ${\sf g}_{{\sf a}{\sf b}}$  takes form of a  3- or 7-dimensional
Euclidean diagonal metric $\delta_{{\sf a}{\sf b}}$.
Introducing
\begin{equation}
J_\mu^{{\sf a}\nu}=\sqrt{\frac{cd}{F^2}}\,F_{\mu}^{{\sf a}\nu},
\label{ImmaginaryUnits}
\end{equation}
equations (\ref{FF'}) take the form
\begin{subequations}\label{FF''}
\begin{equation}
J_{\mu}^{{\sf a}\kappa}J_{\kappa}^{{\sf b}\nu}+J_{\mu}^{{\sf b}\kappa}J_{\kappa}^{{\sf a}\nu}=
-2\,{\delta}^{{\sf a}{\sf b}}\delta_\mu^\nu,
\label{Kaehler-Clifford}
\end{equation}
\begin{equation}
J_{\mu}^{{\sf a}\kappa}J_{\kappa}^{{\sf b}\nu}-J_{\mu}^{{\sf b}\kappa}J_{\kappa}^{{\sf a}\nu}=
2\,{k^{{\sf a}{\sf b}}}_{\sf c}J_{\mu}^{{\sf c}\nu},
\label{algebra}
\end{equation}\end{subequations}
with  ${k^{{\sf a}{\sf b}}}_{\sf c}=\sqrt{\frac{cd}{F^2}}\,{c^{{\sf a}{\sf b}}}_{\sf c}$.
This is enough to exclude $S^7$ (as well as $S^7_3$) as a possible internal space, because there are
not seven roots of unit closing any associative (matrix) algebra. We are left with the sole
$c=3$ case. A direct inspection of the isometry algebra of $S^3$, shows that the three Killing vector
fields parallelizing the 3-sphere and setting ${\sf g}_{{\sf a}{\sf b}}=\delta_{{\sf a}{\sf b}}$,
close the $su(2)$ algebra with structure constants
${c^{{\sf a}{\sf b}}}_{\sf c}=\sqrt{\frac{2|R^\mathrm{in}|}{3}}{\varepsilon^{{\sf a}{\sf b}}}_{\sf c}$.
Taking into account (\ref{F^2 vs R^in}), we therefore obtain
\[{k^{{\sf a}{\sf b}}}_{\sf c}={\varepsilon^{{\sf a}{\sf b}}}_{\sf c}.\]
The rescaled gauge fields $J_{\mu}^{{\sf a}\nu}$ close the quaternionic algebra (\ref{quaternionic algebra}).
To conclude that $g_{\mu\nu}$, $J_{\mu}^{{\sf a}\nu}$ actually define a quaternionic K\"{a}hler
structure of constant quaternionic sectional curvature, we just have to bring together a few formulas scattered
around the paper. By their very definition (\ref{ImmaginaryUnits}), the three complex structures $J_{\mu}^{{\sf a}\nu}$
are isometries of the external space
\begin{subequations}\label{HK'}
\begin{equation}
J_{\mu}^{{\sf a}\kappa}J_{\nu}^{{\sf a}\lambda}g_{\kappa\lambda}=g_{\mu\nu},
\label{Hermitian/anti-Hermitian}
\end{equation}
(no sum over ${\sf a}$), while (\ref{DF=0}) and (\ref{DF}) guarantee the $J_{\mu}^{{\sf a}\nu}$ to be parallel
with respect to the Levi-Civita connection associated to $g_{\mu\nu}$
\begin{equation}
\nabla_\kappa J_{\mu}^{{\sf a}\nu}={\varepsilon_{{\sf b}{\sf c}}}^{\sf a}\theta_\kappa^{\sf b}J_{\mu}^{{\sf c}\nu}
\label{Integrability}
\end{equation}\end{subequations}
with $\theta_\mu^{\sf a}=\sqrt{\frac{2|R^\mathrm{in}|}{2}}A_\mu^{\sf a}$. The compatibility conditions (\ref{HK}) are therefore satisfied.
Eventually, taking again into account (\ref{F^2 vs R^in}), equation (\ref{RiemannEx'}) fixes the external Riemann
tensor to
\begin{subequations}\label{constant k'}
\begin{equation}\begin{array}{l}
R_{\mu\nu\kappa\lambda}=\frac{1}{4}\frac{2|R^\mathrm{in}|}{3}
\left(g_{\mu\lambda}g_{\kappa\nu}-g_{\mu\kappa}g_{\lambda\nu}
+ J_{{\sf a}\mu\lambda}J^{\sf a}_{\nu\kappa}- J_{{\sf a}\mu\kappa}J^{\sf a}_{\nu\lambda}
-2 J_{{\sf a}\mu\nu}J^{\sf a}_{\kappa\lambda}\right),
\label{SpaceFormCurvature'}
\end{array}\end{equation}
while the identity
\begin{equation}\begin{array}{l}
J_{\mu\nu}^{\sf a}=\frac{3}{2|R^\mathrm{in}|}\left(\partial_\mu\theta^{\sf a}_\nu-\partial_\nu\theta^{\sf a}_\mu
-{\varepsilon_{{\sf b}{\sf c}}}^{\sf a}\theta^{\sf b}_\mu \theta^{\sf c}_\nu\right),
\end{array}\label{J as a Gauge Field'}
\end{equation}\end{subequations}
follows from the very definition of $F_{\mu}^{{\sf a}\nu}$ as curvature associated to $A_\mu^{\sf a}$. Thus,
(\ref{constant k}) are satisfied.

\subsection*{Conclusions}
We proved that the equations for conformal flatness (\ref{Weyl=0}) of  arbitrary non-Abelian Kaluza-Klein spaces
are solved by a set of lower dimensional Kaluza-Klein functions $\kappa_{ij}$, $K_{\sf a}^i$, $g_{\mu\nu}$,
$F_{\mu}^{{\sf a}\nu}=\sqrt{\frac{2|R^\mathrm{in}|}{3}}J_{\mu}^{{\sf a}\nu}$  where:
\begin{enumerate}
  \item $\kappa_{ij}$ is the metric of a 3-sphere of arbitrary constant curvature $R^\mathrm{in}$
  \item $K_{\sf a}^i$ are three orthonormal Killing vector fields parallelizing the 3-sphere
  \item $g_{\mu\nu}$, $J_{\mu}^{{\sf a}\nu}$ define a quaternionic K\"{a}hler
  structure of constant quaternionic sectional curvature $\frac{2|R^\mathrm{in}|}{3}$ on the external space
\end{enumerate}
Proceeding in the very same way, it is immediate to prove that also pseudo-quaternionic and para-quaternionic
(corresponding to the internal space $S^3_1$) K\"{a}hler structures of constant pseudo-/para-quaternionic sectional curvature
are solutions.  We suspect this to be the only possible non-trivial solution, but we could not prove the statement.\\
Vice versa, given any $d$-dimensional (pseeudo-/para-)quaternionic K\"ahler manifold of constant (pseeudo-/para-)quaternionic sectional
curvature, (\ref{KKmetric}) defines a $d+3$ dimensional conformally flat manifold. Recalling that a similar statement holds
for (pseudo-/para-)K\"{a}hler manifolds, we conclude that all maximally symmetric special manifolds are in natural correspondence
with conformally flat, possibly non-Abelian, Kaluza-Klein spaces.


\appendix

\section*{Appendices}
\section{Dimensional reduction of the Cotton tensor}\label{A}
The Cotton tensor is the relevant conformal tensor in three dimensions.
However, its vanishing is a necessary|though not sufficient|condition for
conformal flatness in any dimensions. For completeness and because some of
the following identities simplify the analysis of conformal flatness in
arbitrary dimensions, we present the general Cotton reduction formulas.
The Cotton non-trivial property
${{\mathbf C}^J}_{JI}={{\mathbf C}^\nu}_{\nu I}+{{\mathbf C}^j}_{jI}=0$,
makes it convenient to define the quantities
\begin{equation}
{\mathbf C}_I\equiv {{\mathbf C}^\nu}_{\nu I}=-{{\mathbf C}^j}_{jI}.\nonumber
\end{equation}
These are expressed in terms of non-Abelian Kaluza-Klein functions as
\begin{equation}
\begin{array}{l}
{\mathbf C}^{\mu}=\frac{1}{2(D-1)}\hat{\nabla}^\mu\left( c\,
R^\mathrm{ex}+\frac{2d+3c-2}{4}\,F^2\right)
+\frac{1}{4}F_i^{\mu\nu}\hat{\nabla}^\kappa F^i_{\kappa\nu}
\end{array}
\end{equation}
\begin{equation}
\begin{array}{l}
{\mathbf C}_{i}=-\frac{1}{2(D-1)}\nabla_i \left(d\,R^\mathrm{in}-
\frac{3d+4c-4}{4}\,F^2 \right),\label{Ci}
\end{array}
\end{equation}
where here and in the following we take advantage of the fact that only contravariant external and
covariant internal components of higher dimensional tensor properly transform
as lower dimensional tensors. In terms of these quantities Cotton reduction
formulas take the reasonably compact form
\begin{equation}
\begin{array}{lll}
{\mathbf C}^{\mu\nu\kappa}&=&
\frac{2}{d-1}{\mathbf
C}^{[\kappa}g^{\nu]\mu}+C^{\mu\nu\kappa}-\frac{1}{2}\hat{\nabla}_\lambda
(F_i^{\lambda\mu}F^{i\nu\kappa})
+{F_{i\lambda}}^{[\kappa}\hat{\nabla}^{\nu]}F^{i\mu\lambda}\\
&&-\frac{1}{2}F^{i\mu[\nu}\hat{\nabla}_\lambda
F_i^{\kappa]\lambda}-\frac{1}{2(d-1)}F_i^{\lambda[\kappa}g^{\nu]\mu}\hat{\nabla}^\rho
F^i_{\lambda\rho}
+\frac{3}{4(d-1)}g^{\mu[\nu}\hat{D}^{\kappa]}F^2,\label{Cmunukappa}
\end{array}
\end{equation}
\begin{equation}
\begin{array}{lll}
{\mathbf C}_{ijk}&=&-\frac{2}{c-1}{\mathbf
C}_{[k}\kappa_{j]i}+C_{ijk}+\frac{3}{4(c-1)}\nabla_{[k}F^2\eta_{j]i}
-\frac{1}{2}\nabla_{[k}F^2_{j]i},
\label{Cijk}
\end{array}
\end{equation}
\begin{equation}
\begin{array}{lll}
{{\mathbf C}^{\mu\nu}}_k&=&\frac{1}{d}g^{\mu\nu}{\mathbf
C}_k-\frac{1}{2}\hat{\nabla}^\nu \hat{\nabla}_\kappa
F_k^{\mu\kappa} +\frac{1}{2}\nabla_k\left(F^{2\mu\nu}
-\frac{1}{d}F^2g^{\mu\nu} \right)\\
&&+\frac{1}{2}F_k^{\mu\kappa}R_\kappa^\nu
+\frac{1}{2}F_l^{\mu\nu}R^l_{k}+\frac{1}{4}F_k^{\kappa\mu}
F^{2\nu}_\kappa-\frac{1}{8}F^{l\mu\nu}F^2_{lk},
\label{Cmunuk}
\end{array}
\end{equation}
\begin{equation}
\begin{array}{lll}
{{\mathbf C}_{ij}}^\kappa&=&-\frac{1}{c}\kappa_{ij}{\mathbf C}^{\kappa}
+\frac{1}{2}\nabla_j\hat{\nabla}_\lambda F_i^{\kappa\lambda}
-\frac{1}{4}\hat{\nabla}^\kappa\left(F^2_{ij}
-\frac{1}{c}F^2\kappa_{ij}\right)\\
&&-\frac{1}{4} F^\kappa_{i\mu}\hat{\nabla}_\nu
F_j^{\nu\mu}+\frac{1}{4c} F^\kappa_{k\mu}\hat{\nabla}_\nu
F^{k\nu\mu}\kappa_{ij},
\label{Cijkappa}
\end{array}
\end{equation}
\begin{equation}
\begin{array}{lll}
{{\mathbf C}^\mu}_{jk}&=&2{{\mathbf C}_{[kj]}}^\mu,\label{Cmujk}
\end{array}
\end{equation}
\begin{equation}
\begin{array}{lll}
{{\mathbf C}_i}^{\nu\kappa}&=&2{{\mathbf C}^{[\kappa\nu]}}_i\label{Cinukappa}
\end{array}
\end{equation}
where the contracted expressions
$F_k^{\mu\kappa}{F^{k\nu}}_\kappa$,
$F_{i\mu\nu}F_j^{\mu\nu}$,
$F^i_{\mu\nu}F_i^{\mu\nu}$ have been shortened to
$F^{2\mu\nu}$, $F^2_{ij}$ and $F^2$, respectively. Relations
(\ref{Cmunukappa}), (\ref{Cinukappa}) and (\ref{Cijk}),
(\ref{Cmujk}) only hold in $d>1$ and $c>1$ respectively.

\section{Dimensional reduction of the Weyl tensor}\label{B}
The relevant conformal tensor in dimensions greater than three is
the Weyl tensor.
Its property ${{\mathbf C}_{IKJ}}^K={{\mathbf
C}_{I\kappa J}}^\kappa+{{\mathbf C}_{IkJ}}^k=0$  makes it convenient to
introduce the quantities
\begin{equation}
{\mathbf C}_{IJ}\equiv {{\mathbf C}_{I\kappa J}}^\kappa=-{{\mathbf C}_{IkJ}}^k
\hskip0.3cm \mbox{and} \hskip0.3cm
{\mathbf C}\equiv {{\mathbf C}_{\mu}}^\mu=-{{\mathbf C}_{i}}^i.\nonumber
\end{equation}
The lower dimensional scalar ${\mathbf C}$ can be rewritten in terms of $R^\mathrm{ex}$,
$R^\mathrm{in}$ and $F^2$ as
\begin{equation}
\begin{array}{l}
{\mathbf C}=\frac{1}{(D-1)(D-2)}\left[
c(c-1)R^\mathrm{ex}+d(d-1)R^\mathrm{in}+
\frac{(c-1)(2d+3c-2)}{4}F^2\right],
\end{array}
\end{equation}
while the symmetric quantities ${\mathbf C}_{IJ}$ can be rewritten in terms of ${\mathbf C}$
and lower dimensional tensors as
\begin{equation}
\begin{array}{l}
{\mathbf C}^{\mu\nu}=
\frac{1}{d}{\mathbf C}g^{\mu\nu}+\frac{c}{D-2}\left(
R^{\mu\nu}-\frac{1}{d}
R^\mathrm{ex}g^{\mu\nu}\right)+ \frac{d+3c-2}{4(D-2)}
\left(F^{2\mu\nu}-\frac{1}{d}F^2g^{\mu\nu}\right),
\end{array}
\end{equation}
\begin{equation}
\begin{array}{l}
{\mathbf C}_{ij}=-\frac{1}{c}{\mathbf C}\kappa_{ij}
-\frac{d}{D-2}\left( R_{ij}-\frac{1}{c}
R^\mathrm{in}\kappa_{ij}\right)-\frac{c-2}{4(D-2)} \left(F^2_{ij}
-\frac{1}{c}F^2\kappa_{ij}\right),
\end{array}
\end{equation}
\begin{equation}
\begin{array}{l}
{{\mathbf C}^\mu}_j=\frac{c-1}{2(D-2)} \hat{\nabla}_\nu F_j^{\nu\mu}.
\end{array}
\end{equation}
These quantities allow to express Weyl reduction formulas in the reasonably compact
form
\begin{equation}
\begin{array}{lll}
{\mathbf C}^{\mu\nu\kappa\lambda}&=&
\frac{2}{d-2}\left(g^{\mu[\kappa}{\mathbf C}^{\lambda]\nu}
-g^{\nu[\kappa}{\mathbf C}^{\lambda]\mu}\right)
 -\frac{2}{(d-1)(d-2)}{\mathbf
C}g^{\mu[\kappa}g^{\lambda]\nu}+C^{\mu\nu\kappa\lambda}\\
&&+\frac{1}{2}\left(F_i^{\mu\nu}F^{i\kappa\lambda}-F_i^{\mu[\kappa}F^{i\lambda]\nu}\right)
-\frac{3}{2(d-2)}\left(g^{\mu[\kappa} T^{\lambda]\nu}-
g^{\nu[\kappa}T^{\lambda]\mu}\right)
\label{GaussExt}
\end{array}
\end{equation}
with $T^{\mu\nu}=F^{2\mu\nu}-
\frac{1}{2(d-1)}F^2g^{\mu\nu}$,
\begin{equation}
\begin{array}{lll}
{\mathbf C}_{ijkl}&=& -
\frac{2}{c-2}\left(\kappa_{i[k}{\mathbf C}_{l]j} -\kappa_{j[k}{\mathbf
C}_{l]i}\right) -\frac{2}{(c-1)(c-2)}{\mathbf
C}\kappa_{i[k}\kappa_{l]j}+C_{ijkl}
\label{GaussInt}
\end{array}
\end{equation}
\begin{equation}
\begin{array}{lll}
{{\mathbf C}_i}^{\nu\kappa\lambda}&=&
\frac{2}{c-1}g^{\nu[\kappa}{{\mathbf C}^{\lambda]}}_
i-\frac{1}{2}\hat{\nabla}^\nu
F_i^{\kappa\lambda}\label{CodazziExt}
\end{array}
\end{equation}
\begin{equation}
\begin{array}{lll}
{{\mathbf C}^\mu}_{jkl}&=&-\frac{2}{c-1}{{\mathbf C}^\mu}_{[k}\kappa_{l]j}
\label{CodazziInt}
\end{array}
\end{equation}
\begin{equation}
\begin{array}{lll}
{{\mathbf C}^{\mu\nu}}_{kl}&=&\nabla_{k}F_l^{\mu\nu}+
\frac{1}{2}
F_{[k}^{\mu\kappa}{F_{l]\kappa}^\nu} \label{Ricci}
\end{array}
\end{equation}
\begin{equation}
\begin{array}{lll}
{{{{\mathbf C}^{\mu}}_j}^\kappa}_l
&=&{{\mathbf C}^{\mu\kappa}}_{jl}
-\frac{1}{c} {\mathbf C}^{\mu\kappa}\kappa_{jl}+\frac{1}{d}
g^{\mu\kappa}{\mathbf C}_{jl}+\frac{1}{cd}{\mathbf C}
g^{\mu\kappa}\kappa_{jl}\\&&
-\frac{1}{4}F_{(j}^{\mu\nu}{F_{l)\nu}^\kappa}
+\frac{1}{4d}g^{\mu\kappa} F^2_{jl}
+\frac{1}{4c}F^{2\mu\kappa}
\kappa_{jl}-\frac{1}{4cd}F^2g^{\mu\kappa}\kappa_{jl}.\label{6th}
\end{array}
\end{equation}
Relations (\ref{GaussExt}) and (\ref{GaussInt}) only hold in $d>2$ and
$c>2$ respectively. In $d=2$ equation (\ref{GaussExt}) has to be replaced by
${\mathbf C}^{\mu\nu\kappa\lambda}= {\mathbf
C}\,g^{\mu[\kappa} g^{\lambda]\nu}$, while in $c=2$ equation
(\ref{GaussInt}) is substituted by ${\mathbf C}_{ijkl}= {\mathbf
C}\,\kappa_{i[k}\kappa_{l]j}$.

\vskip2.0cm


{\small
\textsc{\begin{minipage}{10.0cm}
Paolo Maraner\\
School of Economics and Management\\
Free University of Bozen/Bolzano\\
via Sernesi 1, Bolzano, 39100, Italy\\
e-mail: pmaraner@unibz.it
\end{minipage}}

\vskip1.0cm

{\small
\textsc{\begin{minipage}{10.0cm}
Jiannis K. Pachos\\
School of Physics and Astronomy\\
University of Leeds\\
Leeds LS2 9JT, UK\\
e-mail: J.K.Pachos@leeds.ac.uk
\end{minipage}}


\begin{thebibliography}{0}

\bibitem{Guralinik&Iorio&Jackiw&Pi03}
G.~Guralinik, A.~Iorio, R.~Jackiw, S.-Y.~Pi,
Dimensionally reduced gravitational Chern-Simons term and its kink,
{\it Ann.\ Phys.\ }{\bf 308} (2003) 222-236.

\bibitem{Grumiller&Jackiw06}
D.~Grumiller, R.~Jackiw,
Kaluza-Klein reduction of conformally flat spaces,
{\it Int.\ J.\ Mod.\ Phys.\ }{\bf D15} (2006) 2075-2094.

\bibitem{Jackiw07}
R.~Jackiw,
Dimensional reduction of conformal tensors and Einstein-Weyl spaces,
{\it SIGMA}{\bf 3} (2007) 91-97.

\bibitem{Grumiller&Jackiw07}
D.~Grumiller, R.~Jackiw,
Einstein-Weyl from Kaluza-Klein,
{\it Phys.\ Lett.\ }{\bf A372} (2007) 2547-2551.


\bibitem{Maraner&Pachos09}
P.~Maraner, J.~K.~Pachos, Conformally flat Kaluza-Klein spaces,
pseudo-/para-complex space forms and generalized gravitational kinks,
{\it J.\ Geom.\ Phys.\ } {\bf 59} (2009) 1314-1325.

\bibitem{Bryant93} R.~L.~Bryant,
Classical, exceptional, and exotic holonomies: a status report,
in Actes de la Table Ronde de G\'{e}om\'{e}trie Diff\'{e}rentielle en l'Honneur de Marcel Berger,
Soc.\ Math.\ France (1993) 93-166.

\bibitem{Yano64}
K.~Yano,
\textsc{differential geometry on complex and almost complex spaces},
Pergamon, 1965.

\bibitem{Barros&Romero82}
M.~Barros, A.~Romero,
Indefinite K\"ahler manifolds,
{\it Math.\ Ann.\ }{\bf 261} (1982) 55-62.

\bibitem{Cruceanu&Fortuny&Gadea96}
V.~Cruceanu, P.~Fortuny, P.~M.~Gadea,
A survey of paracomplex geometry,
{\it Rocky Mountain J.\ Math.\ }{\bf 26}, n.1 (1996)
83-115.


\bibitem{Alekseevskii68}
A.~D.~Alekseevskii,
Riemannian spaces with exceprional holonomy groups,
{\it Funkcional.\ Anal.\ i Prilo\v{z}en  }{\bf 2} (1968) 1-10.

\bibitem{Ishihara74}
S.~Ishihara,
Quaternion K\"ahlerian Manifolds,
{\it J.\ Differential Geometry }{\bf 9} (1974) 483-500.

\bibitem{Salomon82}
S.~Salomon,
Quaternionic K\"ahler Manifolds,
{\it Inv.\ Math.\ }{\bf 67} (1982) 143-171.

\bibitem{Garcia-Rio&Matsushita&Vazquez-Lorenzo01}
E.~Garc\'ia-R\'io, Y.~Matsushita, R.~V\'azquez-Lorenzo,
Paraquaternionic K\"ahler Manifolds,
{\it Rocky Mountain J.\ Math.\ }{\bf 31} (2001) 237-260.

\bibitem{Gadea&AmilibiaMontesinos89}
P.~M.~Gadea, A.~Montesinos~Amilibia,
Spaces of constant para-holomorphic sectional curvature,
{\it Pacific J.\ Math.\ }{\bf 136}, (1989) 85-101.

\bibitem{Gadea&MunozMasque92}
P.~M.~Gadea, J.~Mu\~{n}oz~Masqu\'{e},
Classification of homogeneous parakaehlerian space forms,
{\it Nova J.\ Alg.\ Geo.\ }{\bf 1}, (1992) 111-124.

\bibitem{Perez&Santos82}
J.~D.~P\'erez, F.~G.~Santos,
Indefinite Quaternion Space Forms,
{\it Annali di matematica pura ed applicata }{\bf 132} (1982) 383-398.

\bibitem{Blazic92}
N.~Bla\v{z}i\'{c},
Paraquaternionic projective space and pseudo-Riemannian geometry,
{\it Publ.\ Math.\ (Beograd)} {\bf 60}, (1996) 101-107.

\bibitem{KK}
R.~Coqueraux, A.~Jadczyk,
\textsc{riemannian geoemtry, fiber bundles, kaluza-klein theories and all that},
World Scientific, 1988.

\bibitem{O'Neill83}
B.~O'Neill,
\textsc{semi-riemannian geometry with application to relativity},
Academic Press, 1983.

\bibitem{D'Atri&Nickerson68}
J.~E.~D'Atri, H.~K.~Nickerson,
The existence of special orthonormal frames,
{\it J.\ Differential Geometry }{\bf 2} (1968) 393-409.

\bibitem{Wolf72}
J.~A.~Wolf,
On the geometry and classification of absolute parallelism I \& II,
{\it J.\ Differential Geometry }{\bf 6} (1972) 317-342, {\bf 7} (1972) 19-44.



\end{thebibliography}
\end{document}